\documentclass[10pt,journal,cspaper,compsoc]{IEEEtran}

\ifCLASSOPTIONcompsoc
\else
\fi
\ifCLASSINFOpdf
\else
\fi
\newcommand{\subparagraph}{}

\usepackage[square, comma, sort&compress, numbers]{natbib}
\usepackage{graphicx}
\usepackage{array}
\newcommand{\ApproxSign}{\raise.17ex\hbox{$\scriptstyle\sim$}}
\usepackage[table]{xcolor}
\usepackage{multirow}
\usepackage{titlesec}

\let\labelindent\relax
\usepackage{enumitem}
\setlist[enumerate]{leftmargin=*}

\usepackage{textcomp}
\usepackage[normalem]{ulem}
\usepackage{url}
\DeclareUrlCommand\ULurl{%
  \renewcommand\UrlLeft{\bgroup}%
  \renewcommand\UrlRight{\egroup}}
\usepackage[linesnumbered,lined,ruled,commentsnumbered]{algorithm2e}
\usepackage{amsmath}
\usepackage[table]{xcolor}
\usepackage{color}

\usepackage{amssymb}
\usepackage{pifont}
\newcommand{\cmark}{\ding{51}\ }%
\newcommand{\xmark}{\ding{55}\ }%
\newcolumntype{L}{>{\centering\arraybackslash}m{3.5cm}}
\newcolumntype{G}{>{\centering\arraybackslash}m{10.5cm}}
\begin{document}

%
\title{A Survey of Techniques for Improving Security of GPUs}

\author{Sparsh Mittal, S.B. Abhinaya, Manish Reddy and Irfan Ali
 
\IEEEcompsocitemizethanks{\IEEEcompsocthanksitem S. Mittal, M. Reddy and I. Ali are with IIT Hyderabad, India.  Abhinaya is with NIT Trichy and she contributed to this paper while working as an intern at IIT Hyderabad. Support for this work was provided by Science and Engineering Research Board (SERB),
India, award number ECR/2017/000622. 
\protect\\
}
\thanks{}}


%
%

\markboth{}%
{author : A Survey }

\onecolumn
\IEEEcompsoctitleabstractindextext{%

\begin{abstract}
Graphics processing unit (GPU), although a powerful performance-booster, also has many security vulnerabilities. Due to these, the GPU can act as a safe-haven for stealthy malware and the weakest `link' in the security `chain'.  In this paper, we present a survey of techniques                                                                  for analyzing and improving GPU security. We classify the works on key attributes to highlight their similarities and differences. More than informing users and researchers about GPU security techniques, this survey aims to increase their awareness about GPU security vulnerabilities and potential countermeasures.

\end{abstract}

\begin{IEEEkeywords}
GPU, security, review,  side-channel, covert channel, encryption, buffer overflow, malware, information leakage.
\end{IEEEkeywords}}

\maketitle

\IEEEdisplaynotcompsoctitleabstractindextext

%
\IEEEpeerreviewmaketitle

%

%
%
%

%


\section{Introduction}\label{sec:introduction}

The computing industry is currently at an interesting inflection point. GPUs, which were originally used for a narrow range of graphics (e.g., video-gaming) applications, are now spreading their wings to a broad spectrum of compute-intensive and mission-critical applications, most notably, cryptography, finance, health, space and defense. After passing the initial `rounds'   of scrutiny on the metrics of performance and energy \cite{mittal2015gpupowermgmtsurvey,mittal2015cpugpusurvey},  it is time that GPUs face and pass the test on the metric of security, which is especially crucial in mission-critical applications.

In fact, a recent incident has strongly highlighted the need and even urgency of  improving GPU security. A malicious person hid a bitcoin miner in ESEA (a video game service) software \cite{eseaBitCoin}. This miner used the GPUs in users' machines for mining bitcoin without their knowledge. The miner overheated and harmed the machines by overloading the GPUs. Thus, the malicious person earned cryptocurrency at the expense of the users' resources. This incident shows that the community can no longer afford to ignore GPU security considerations. 

While a large part of the conventional wisdom gained from CPU security research is also applicable to GPU security, both \textit{attacking} and \textit{securing} GPU present challenges of their own. Many attacks exploit the correlation between an event and its impact such as the change in latency, power consumption or number of memory accesses. However, due to its massively-parallel architecture and undocumented management policies, isolating individual events and their impact is generally not feasible. These very reasons along with the fast-evolving architecture of GPUs, also make it difficult to design effective security solutions for them \cite{zhu2017understanding}. These factors demand a careful study of GPU security. Several recent works reveal security vulnerabilities of GPUs along with their countermeasures.

\begin{figure} [h]
\centering
\includegraphics[scale=0.40]{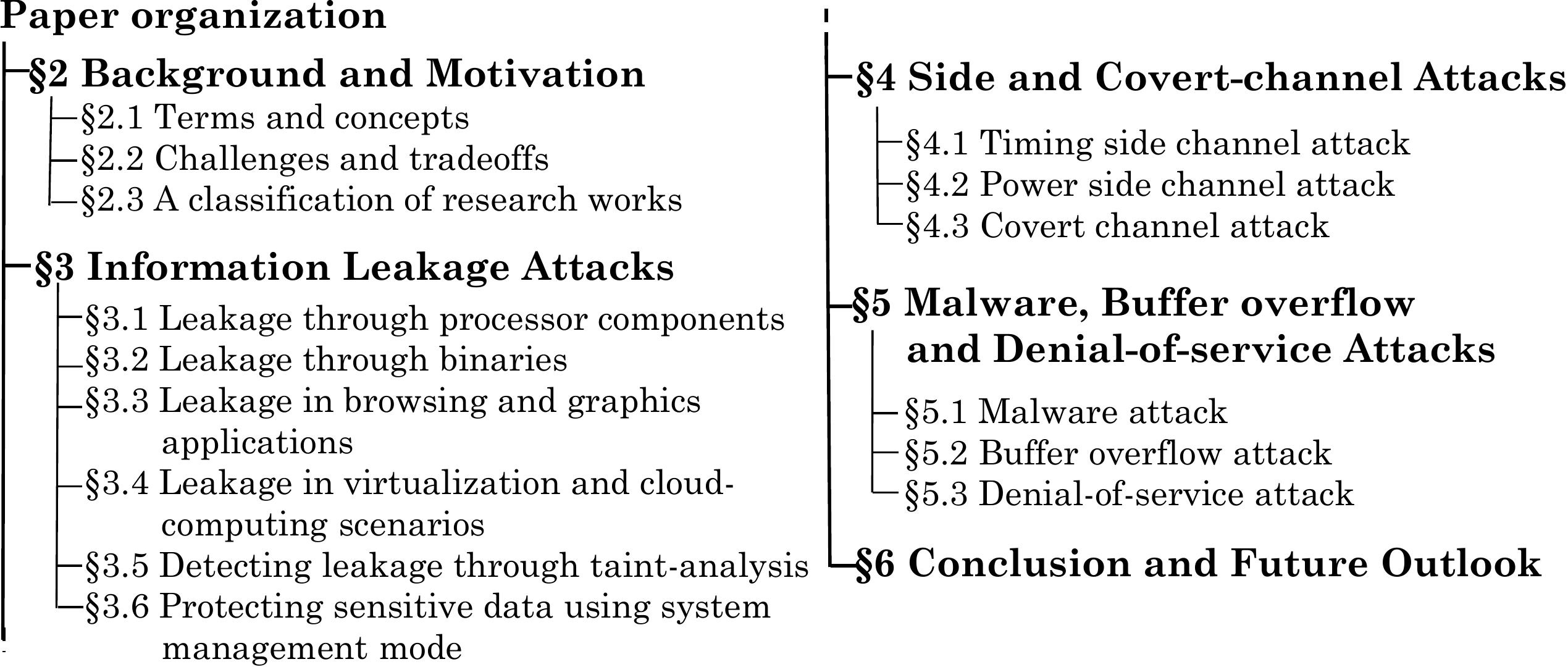}
\caption{Organization of the paper }\label{fig:overview}
\end{figure}

\textbf{Contributions:} In this paper, we present a survey of techniques for analyzing and improving the security of GPUs. Figure \ref{fig:overview} presents an overview of the paper. Section \ref{sec:background} first presents a brief background on important concepts and terms. Then, it highlights the need for improving GPU security and tradeoffs involved in protecting them.
After this, it presents a classification of research works and also summarizes their key ideas. In security jargon, attacks are divided into two types: ``passive attacks'' which leak system-information but do not the change the system, and ``active attacks'' which change the data or operation of the system. We review passive attacks in Sections \ref{sec:dataleakage} and \ref{sec:sidechannel} and active attacks in Section \ref{sec:malwareOverflowDoS}. Specifically, Section \ref{sec:dataleakage} reviews attacks for leaking sensitive information. Section \ref{sec:sidechannel} discusses side and covert-channel attacks. Section \ref{sec:malwareOverflowDoS} discusses GPU malware, buffer-overflow and DoS\footnote{The following acronyms are frequently used in this paper: advanced encryption standard (AES),  address space layout randomization (ASLR),  bandwidth (BW), covert channel (CC),  compute unified device architecture (CUDA), direct memory access (DMA), denial of service (DoS),   error correcting code (ECC),    functional unit (FU), global/local/shared memory (GlM/LoM/ShM), input output memory management unit (IOMMU), input output (I/O),   memory management unit (MMU),  peripheral component interconnect (PCI), parallel thread execution (PTX),  single instruction multiple thread (SIMT),  streaming multiprocessor (SM),  system management mode (SMM), shared virtual memory (SVM), virtual machine (VM).}  attacks. In these sections, we discuss a work under single category only, even though many of the works span across the boundaries. 

\textbf{Scope:} For the sake of a clear presentation, we limit the scope of this paper as follows. We include only those techniques which have been proposed and evaluated in the context of GPUs, even though some of the security techniques proposed in the context of CPUs may also be applicable for GPUs. We do not include works which use GPU merely for accelerating a cryptography algorithm, but those that seek to leak the encryption key. We include works that attack GPU or run attack-code/malware on GPU. We believe that this survey will underscore the need of designing GPUs with security as the first principle rather than retrofitting for it.

\section{Background and Motivation}\label{sec:background}
\subsection{Terms and concepts}\label{sec:terms}

We now introduce some concepts which will be used throughout this paper.  We refer the reader to previous works  for a detailed background on  
GPU architecture \cite{mittal2016SurveyGPURF,mittal2014surveyGPUcache,mittal2017design} and AES encryption algorithm \cite{kadam2018rcoal}.

\textbf{{\tt cudaContext}}: It is a data type which stores device configuration during runtime such as error codes, loaded modules containing device code and allocated device memory. These parameters are useful for controlling the device. 
  
\textbf{{\tt cl\_mem:}} In OpenCL execution model, a memory object is a handle to a portion of  global memory. {\tt cl\_mem} is an object type used for describing the memory objects \cite{opencl_clmem}.

\textbf{Shader:} A shader refers to a program which processes a particular stage of a graphics engine for calculating rendering effects. For example, a fragment shader is a shader stage which processes fragments into multiple colors and a depth value. The vertex shader is another stage which processes individual vertices. 

\textbf{Memory access coalescing:} It refers to merging of memory requests of different threads of a single warp into as few cache lines as possible. 

\textbf{Soft and hard booting:} In soft (or warm)  rebooting, the system is restarted without turning off the power supply to the system. In hard (or cold) rebooting, the power to the system is first turned-off (i.e., shut-down) and then turned-on which results in rebooting the system.

\textbf{SMM:} SMM is a special operating mode in x86 CPUs which is used for performing tasks related to system-control, e.g., hardware management \cite{kim2016demand}. SMM can provision a memory portion which can be accessed only by SMM and not hypervisor or OS kernel. Further, in SMM, only a trusted program, namely ``system management interrupt'' (SMI) can run. All other tasks, including any malicious task, are suspended and they cannot interrupt the SMI program. Note that a hypervisor is a firmware which launches and runs virtual machines.

\textbf{DMA:} DMA allows certain devices to directly access main memory without requiring the intervention of CPU. 
  
\textbf{Buffer overflow:} Buffer overflow refers to writing data outside the boundary of the buffer to the adjacent locations. It can lead to program crashes, data corruption, and security breaches. For example, stack overflow by a thread can impact execution of other threads by overwriting other memory spaces.  

\textbf{Canary:} In memory management, a canary is the memory location which does not store useful data, and is generally placed just before the return address of a function. An adversary generally attempts to cause a buffer overflow to overwrite the return address for redirecting the execution to a malicious code. However, use of a canary ensures that the buffer overflow also overwrites the canary value. Then, before using the return address, the canary can be  checked. If it has been changed, overwriting of the return address can be easily detected. 

\textbf{Covert and side-channel:} A channel refers to a medium through which sensitive data is leaked. If the channel is hidden, it is termed as ``covert channel''. A covert channel is created intentionally and is not otherwise meant for communication. The adversary tries to conceal its existence from the victim.   A ``side-channel'' is created incidentally, where the adversary gets sensitive information from the characteristics of the system's operation \cite{wang2006covert}. For example,  if the timing/power values of the encryption algorithm depend on the key, then, based on the timing/power measurements, an adversary can guess the key used in encryption.  In a side-channel, there is no communication, but only leakage of sensitive information through the side-channel.

\textbf{Denial-of-service attack:} In a DoS attack, the adversary tries to make the device unavailable by temporarily or permanently hampering its services. This may be done by overloading the system with useless requests which prohibits handling of genuine requests from a benign user. Due to this, a DoS attack can be detected, which is different from other attacks, such as side-channel attack, where the system is generally not harmed and hence, no evidence of the attack remains.

\subsection{Challenges and tradeoffs}\label{sec:challengesandtradeoffs}
Protecting GPUs, although important, presents several challenges, as we discuss below.  

\textbf{Limitations of CPU-based security solutions:} After launching the program on GPU (device), CPU (host) remains isolated and thus, it cannot monitor the execution of GPU. Hence, security mechanisms proposed on CPUs, such as a CPU taint-tracking scheme may not work for GPUs. For example, they may not detect a GPU-resident malware and thus, an attacker can use GPU as the polymorphic malware extractor whereby  the host can load the compressed/encrypted code on GPU and then call a GPU kernel to quickly unpack/decrypt the code \cite{patterson2013vulnerability}. Also, the GPU can be used for launching a brute force attack to crack the passwords. Similarly, since a sharp increase in GPU load is likely to go undetected more easily  compared to that in CPU load, a GPU malware is more stealthy. In fact, given the close interaction between GPU and the host, a compromised GPU also threatens the security of  host processor. 

\textbf{Lack of documentation and open-source tools:} Existing GPU vendors take ``security-through-obscurity approach'' for securing GPUs. As such they do not reveal complete information about GPU microarchitecture and thus, most of this information is obtained  through reverse engineering attempts only \cite{black2010cubar,envytools,naghibijouybari2017constructing}. Lack of official documentation allows GPU vendors to introduce architectural changes for boosting performance, even at the cost of jeopardizing GPU security. The scarcely-available documentation on GPU discusses only performance, and not security-related issues. 

For example, GPU binary utilizes closed-source assembly language which cannot be inspected by existing anti-virus tools. Similarly, GPU vendors do not define/document how the deallocated memory is handled  \cite{pietro2016cuda}.  Further, in GPUs, the separation of responsibilities between hardware/software is not clear and has been changing over GPU generations \cite{zhu2017understanding}. Due to this, reasoning about security guarantees becomes difficult.

\textbf{Lack of data erasure:} GPU hardware/drivers do not erase their memories and thus, in ShM, LoM, GlM and registers, data persists after deallocation \cite{pietro2016cuda,patterson2013vulnerability}.  
Further, while most kinds of memories can be deleted by users, GPUs do not allow users to erase some kinds of memories which store sensitive data such as kernel codes, constant data and call-by-value arguments \cite{lee2014stealing}. Also, since the LoM and registers are compiler-managed, erasing them is not straightforward.  By exploiting this, an adversary can leak sensitive information.  For example, to leak information from the register file, an adversary can write a kernel with the same occupancy and thread-block size as the victim kernel. This ensures similar, predictable partitioning of register file. Then, the malicious kernel can be coded in a way to read the target registers.

\textbf{Increasing reliance on GPUs:} GPUs are increasingly being used for accelerating a wide variety of applications, such as browsers, document processors (e.g., Adobe reader, Microsoft Office, Libre Office), scientific computing (MATLAB) and image processing tools, etc. For instance, WebGL allows browsers to utilize GPUs for accelerating the rendering of webpages. In fact, Google Chrome provides an option for using GPU to render the  entire webpage and not merely WebGL content \cite{chromeGPU}. However, this can be leveraged to launch DoS attack remotely by somehow making a user open a malicious website which overloads users' GPUs.  Also notice that in these graphics applications, the ability to read/write GPU memory can allow an adversary to change the contents displayed on user's screen. 

Similarly, in mission-critical applications where GPUs store and process sensitive data, an attack on them can have huge financial and social consequences. For example, Zhou et al. \cite{zhou2017vulnerable} demonstrate extraction of credit card numbers and email contents from remanent data in GPU memory. Also, in companies, a malicious insider may access classified documents  which were opened on a shared GPU by an authorized user.

\textbf{Characteristics of GPU architecture and usage model:} In GPUs, the presence of multiple memories with different access rights and lifetimes complicates their security solutions and mandates individual security solutions for them. Also, since even non-privileged users can run GPU programs, a large number of users can attack or exploit GPU. Further, once the data is copied from the host memory to device memory, it is managed by the GPU driver and not the host processor/OS. However, GPU drivers may not be as rigorously evaluated from the security perspective as the existing operating systems.

\textbf{Vulnerability in virtualization and cloud scenarios:} As the computational capabilities of GPU increase, a single GPU is increasingly being shared among  multiple users. As such, major cloud services provide GPU computing platforms such as Amazon web services, Google cloud platform and IBM cloud \cite{naghibijouybari2017constructing}.  However, different users in the cloud computing scenario may not trust other. For example, an adversary can rent a GPU-based VM and leak information of users using other VMs on the same system via GPU memory. Clearly, with GPU virtualization approach, the risks of information-leakage is even higher than that with native execution.

\textbf{Vulnerability due to buffer overflows:} Conventionally, GPU and CPU memory were separated due to which corruption of CPU memory by GPU was difficult. Also, GPU applications did not typically perform tasks such as pointer-dereferencing, and hence, buffer corruption did not cause much harm \cite{erb2017dynamic}. Further, since GPU memory was sparsely allocated, buffer overflow did not lead to overwriting useful data. However, as recent GPUs share virtual and even physical memory with CPUs \cite{hsaRef}, buffer overflows in GPU can also lead to security and correctness problems \cite{erb2017dynamic}. Existing tools to detect GPU buffer overflow, such as Oclgrind may incur latency overheads of up to 300 times \cite{price2015oclgrind}. 

\textbf{Challenges in attacking GPUs: } While securing GPUs is challenging as mentioned above, attacking GPUs also presents challenges. By virtue of its massively parallel architecture, GPUs can simultaneously perform multiple encryptions and hence, the timing of individual encryptions cannot be measured. In fact, the encryption with the highest latency dominates the overall latency.  Further, since only one {\tt cudaContext} can run on GPU at any time, a data-leakage attack can obtain only the final snapshot of the previous process \cite{pietro2016cuda}. Similarly, in a cloud environment, both the cloud and GPU architectures offer layers of obscurity which makes it difficult to launch an attack on GPUs. 
    
The techniques discussed in the rest of this article propose strategies for overcoming these challenges.

\subsection{A classification of research works}\label{sec:classification}
 
Table \ref{tab:classification1} classifies the research works based on attack type and the GPU component being studied. It is evident that nearly all important GPU components are vulnerable to attacks. Table \ref{tab:classification2} organizes the works on the basis of evaluation platform and programming language. Clearly, the GPUs from all major vendors are vulnerable to attacks.

\begin{table}[htbp]
\centering
\caption{A classification based on attack type and GPU component}
\label{tab:classification1}
\begin{tabular}{|p{3.0cm}|p{13cm}|}\hline
\multicolumn{1}{|c}{Category}   & \multicolumn{1}{|c|}{References} \\\hline\hline
\multicolumn{2}{|c|}{Type of attack or security vulnerability}   \\\hline
Information leakage &  \cite{patterson2013vulnerability,maurice2014confidentiality,bellekens2016strategies,pietro2016cuda,jiang2017novel,jiang2016complete,zhou2017vulnerable} \\\hline
Side-channel attack & timing  \cite{jiang2017novel,kadam2018rcoal,jiang2016complete}, power \cite{luo2015side} \\\hline
Covert-channel attack & \cite{naghibijouybari2017constructing} \\\hline
Malware &  \cite{vasiliadis2015gpu,zhu2017understanding,hayes2017gpu,ladakis2013you,naghibijouybari2017constructing} \\\hline
Buffer overflow &  \cite{miele2016buffer,erb2017dynamic,di2016study} \\\hline
Denial of service &  \cite{patterson2013vulnerability,balzarotti2015impact} \\\hline

\multicolumn{2}{|c|}{GPU components}   \\\hline
Global memory & information leakage  \cite{pietro2016cuda,maurice2014confidentiality,lombardisecure,lee2014stealing,hayes2017gpu,zhang2015forensically,bellekens2015data},  side-channel  \cite{jiang2016complete,kadam2018rcoal}, covert channel \cite{naghibijouybari2017constructing}, buffer overflow \cite{erb2017dynamic,di2016study} \\\hline

Shared memory & information-leakage \cite{lee2014stealing,bellekens2015data,pietro2016cuda,lombardisecure,hayes2017gpu}, side-channel \cite{jiang2017novel,jiang2016complete}, covert channel \cite{naghibijouybari2017constructing},  buffer overflow \cite{di2016study} \\\hline

Local memory & information-leakage \cite{pietro2016cuda,lombardisecure,lee2014stealing,hayes2017gpu} \\\hline
Texture memory & information-leakage \cite{bellekens2015data} \\\hline
Registers & information leakage \cite{pietro2016cuda,lombardisecure,hayes2017gpu}, side-channel \cite{jiang2016complete}
\\\hline
Cache & information-leakage \cite{lee2014stealing}, covert channel \cite{naghibijouybari2017constructing} \\\hline
Functional unit & covert-channel \cite{naghibijouybari2017constructing}  \\\hline
\multicolumn{2}{|c|}{Application domain or context where attack is shown}   \\\hline

Virtualization/cloud-computing &  \cite{maurice2014confidentiality,zhou2017vulnerable,lombardisecure} \\\hline
Browsing/graphics applications &  \cite{lee2014stealing,zhou2017vulnerable,zhang2015forensically,danisevskis2013dark,patterson2013vulnerability,hayes2017gpu} \\\hline

Encryption algorithm & AES   \cite{kadam2018rcoal,pietro2016cuda,jiang2017novel,jiang2016complete,luo2015side,kim2016demand}, RSA \cite{kim2016demand} \\\hline
Keylogging & \cite{ladakis2013you} \\\hline

\end{tabular}
\end{table}
 
\begin{table}[htbp]
\centering
\caption{A classification based on  evaluation platform and programming language}\label{tab:classification2}
\begin{tabular}{|p{3.0cm}|p{6cm}|}\hline
\multicolumn{1}{|c}{Category}   & \multicolumn{1}{|c|}{References} \\\hline\hline

\multicolumn{2}{|c|}{Overall goal}   \\\hline
Security & all \\\hline
Performance & \cite{hayes2017gpu,kadam2018rcoal} \\\hline
\multicolumn{2}{|c|}{Evaluation platform}   \\\hline
GPU simulator &  \cite{kadam2018rcoal} \\\hline
real GPU & nearly all others \\\hline
\multicolumn{2}{|c|}{GPU make}   \\\hline
Intel GPU &  \cite{patterson2013vulnerability,di2016study,balzarotti2015impact,miele2016buffer,lombardisecure} \\\hline
AMD GPU &  \cite{lee2014stealing,zhou2017vulnerable,jiang2016complete} \\\hline
ARM GPU & \cite{danisevskis2013dark} \\\hline
NVIDIA GPU & nearly all \\\hline
\multicolumn{2}{|c|}{Programming language}   \\\hline
OpenCL &  \cite{erb2017dynamic,lee2014stealing,jiang2016complete,bellekens2016strategies,zhou2017vulnerable,hayes2017gpu} \\\hline
CUDA &  \cite{lombardisecure,jiang2016complete,kim2016demand,zhang2015forensically,luo2015side,naghibijouybari2017constructing,patterson2013vulnerability,pietro2016cuda,zhu2017understanding,jiang2017novel,zhou2017vulnerable,lee2014stealing,ladakis2013you,maurice2014confidentiality,miele2016buffer,bellekens2015data,di2016study,bellekens2016strategies,pietro2016cuda,hayes2017gpu,kadam2018rcoal} \\\hline
\multicolumn{2}{|c|}{Other features}   \\\hline
Binary instrumentation &  \cite{hayes2017gpu,erb2017dynamic} \\\hline
Fast Fourier transform & \cite{zhou2017vulnerable} \\\hline
\end{tabular}
\end{table}

 \section{Information-leakage Attacks}\label{sec:dataleakage}
In this section, we discuss techniques which demonstrate information-leakage through various processor components such as ShM, GlM and texture memory (Section \ref{sec:leakcomponents}), and through GPU binaries (Section \ref{sec:leakagebinaries}). 
Also, we discuss leakage in different contexts, such as web-browsing/graphics applications (Section \ref{sec:leakagewebbrowser}) and virtualized systems (Section \ref{sec:leakagevirtualized}). Finally, we discuss use of taint-analysis technique for tracking the flow of sensitive data (Section \ref{sec:taintanalysis}) and use of SMM for protecting sensitive data (Section \ref{sec:kimSMM}).

\subsection{Leakage through processor components}\label{sec:leakcomponents}   
  
\begin{figure} [h]
\centering
\includegraphics[scale=0.40]{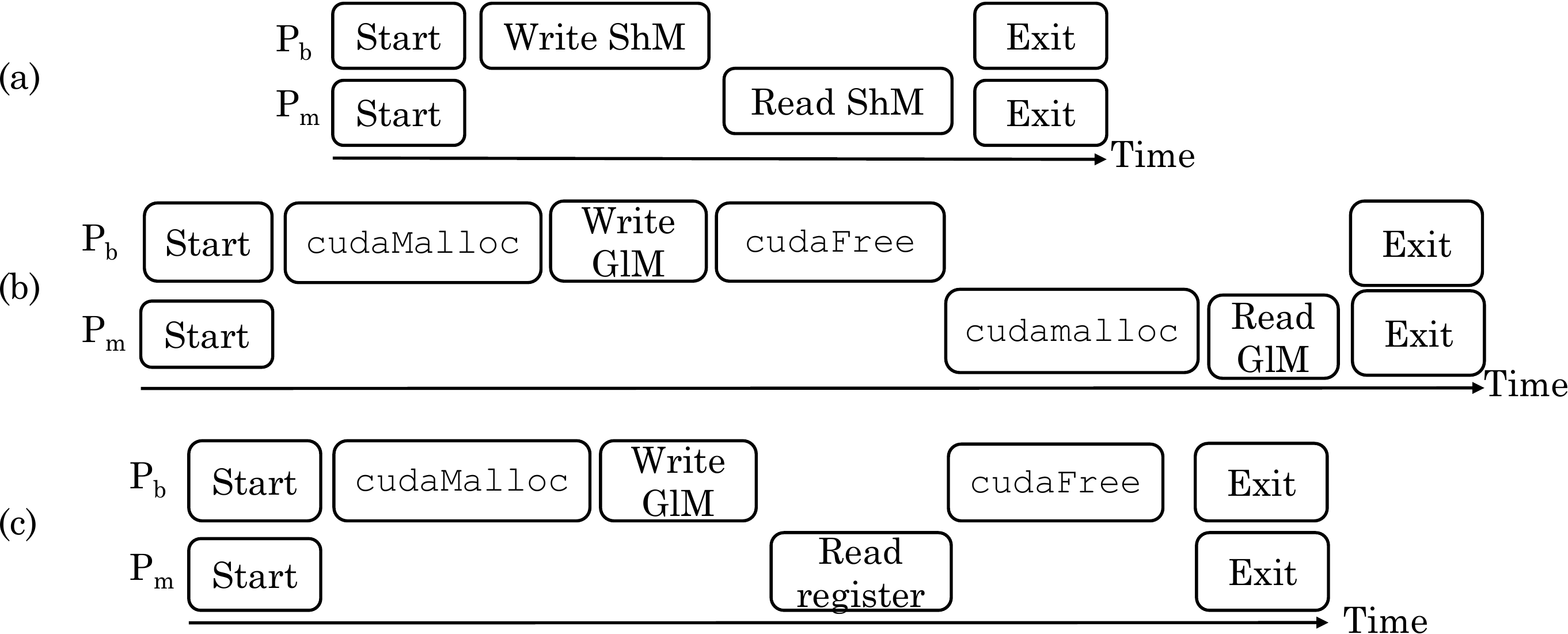}
\caption{The sequences for causing leakage on (a) ShM (b) GlM and (c) registers \cite{pietro2016cuda} }\label{fig:sequenceForLeakage}
\end{figure}

Pietro et al. \cite{pietro2016cuda} show information leakage in ShM, GlM and registers by using CUDA memory (de)allocation commands. Let $P_b$  and $P_m$ refer to a benign and a malicious process, respectively. For attacking the ShM, assume that $P_b$ runs a kernel which copies a vector from GlM to ShM. This kernel is run $Q$ times. The size of the vector is equal to that of ShM. $P_m$ runs a kernel which reads ShM and this kernel is also run for $Q$ times.   Specifically, $P_m$ allocates a vector in ShM which is of the same size as ShM. Then, it copies data from this vector to a vector in GlM. Using this, $P_m$ can read the entire data written by $P_b$ in ShM. The only requirement is that $P_m$ should be scheduled before $P_b$ terminates. The attack sequence is summarized in Figure \ref{fig:sequenceForLeakage}(a). They also observe that, before the host executes {\tt exit()} function, the ShM is zeroed. 
    
For attacking global memory, assume that the $P_b$ allocates 4 vectors (V1 to V4) of equal size using {\tt cudaMalloc}. Then, it initializes V1 and V2 and copies them to V3 and V4, respectively. Then $P_b$ terminates and $P_m$ is scheduled. Using Unix sockets, correct synchronization is maintained between them. $P_m$ also allocates four vectors of the same size as done by $P_b$. However, instead of initializing, $P_m$ executes the same kernel code as executed by $P_b$ and copies V3 and V4 to host memory. They observe that $P_m$ finds the same data as written by $P_b$. Thus, the entire data written on global memory can be leaked in a deterministic manner, as long as $P_m$ allocates the same amount of memory as released by $P_b$. Figure \ref{fig:sequenceForLeakage}(b) summarizes the sequence for causing leakage in GlM. 

For leaking data through the register, they exploit ``register spilling'' mechanism . A process can reserve more registers than those available on the GPU. Variables which cannot be placed in the register are placed in GlM and this is termed as register spilling. An attacker can exploit this to access GlM reserved for other CUDA contexts, even when the benign process owns them and has not freed them using {\tt cudaFree}. This makes this attack very dangerous. As for the attack procedure, $P_b$ writes to GlM multiple times and $P_m$ attempts to read the memory allocated to $P_b$. $P_m$ reserves a fixed number of registers and copies their content to host memory. They executed $P_b$ and $P_m$ simultaneously and observed that $P_m$ could read the memory reserved by $P_b$ before it is released. Thus, the attacker can bypass runtime access primitives.  However, this attack does not allow interfering with the computations performed by the benign process. Also, this attack was successful on Kepler GPU but not on Fermi GPU. The event sequence for causing leakage in registers is summarized in Figure \ref{fig:sequenceForLeakage}(c). Table \ref{tab:summaryOfCUDALeaks} summarizes the three attacks proposed by them. 

\begin{table}[htbp]
  \centering 
  \caption{Characteristics of attacks on ShM, GlM and registers \cite{pietro2016cuda}}
    \begin{tabular}{|l|l|p{9cm}|} \hline
          & Leakage & Conditions \\ \hline
    Shared Memory & Complete & $P_m$ should be scheduled before $P_b$ terminates  
    \\ \hline
    Global Memory & Complete & $P_{b}$ has terminated and $P_{m}$ allocates the same amount of memory as $P_{b}$ \\ \hline
    Registers & Partial & None \\ \hline
    \end{tabular}%
  \label{tab:summaryOfCUDALeaks}%
\end{table}%

As for mitigation of attacks on GlM and ShM, they propose the use of data-shredding (i.e., zeroing).  For thwarting register-attacks, the GPU driver should (1) forbid spilling to memory locations in GlM which are already reserved by other processes and (2) reset the locations of spilled registers after releasing them.

Bellekens et al. \cite{bellekens2015data} discuss ways in which remanent sensitive data may be leaked from the global, shared and texture memory of a GPU. For GlM, a benign user transfers data from the host to the GPU using {\tt cudaMalloc} and {\tt cudaMemcpy}. Then, the adversary executes a program which allocates same or larger amount of memory using {\tt cudaMalloc}. Then, it dumps the left-over data in GPU memory to the CPU. By allocating the memory, the adversary can partially or fully recover the data in GPU memory.

For leaking data from ShM, they send the data to GlM and then store it in ShM. Then, the adversary runs a program which requires a maximum size uninitialized array in ShM. Using standard I/O, the adversary can get the data of ShM or dump it to global memory. For leaking data from the texture memory, they first transfer the data from CPU and bind it into texture memory using {\tt cudaBindTextureToArray}. Then, the adversary can use the same approach as that used for ShM to leak the data. 

They note that different grades of GPUs, viz., server, consumer and mobile grades are vulnerable to such leaks. Further, data remanence can be avoided only by using ``hard reboot'' since the ``GPU reset'' mechanism does not prevent such leaks. As for software schemes for mitigating data leaks, they propose writing zero or misleading data during initialization. This can be performed using a separate thread. Also, along with the benign process, a secondary process may be run on the GPU. In case of abrupt termination of the benign process, the secondary process can immediately erase the data to avoid data-leak.

Patterson \cite{patterson2013vulnerability} notes that unlike CPUs, GPUs do not implement ASLR (``address space layout randomization'') or virtual memory. Hence, on GPUs, pointer allocations repeatedly return the same addresses and thus, the data-address can be determined. In the absence of virtual memory, logical addresses of different processes access the same physical memory \cite{mittal2017SurveyTLB}. Since GPUs do not erase the memory, the data persists in memory even after program termination. This attack can be extended to find specific data types. They also note that this attack succeeded even across different users and login sessions. 
    
\subsection{Leakage through binaries}\label{sec:leakagebinaries}

Bellekens et al. \cite{bellekens2016strategies} note that CUDA compilers generate ELF binaries which have high-level assembly PTX code. Note that ELF is a file format for shared binaries, object code and executable files.  This information about assembly code may be used for malicious purposes or reverse engineering. They show that by modifying and compiling PTX code in a just-in-time manner, even simple dynamic analysis can reveal details about the source-code. Combining this with debugging tools allows performing both dynamic and static analysis of the binaries. To thwart these attacks, they suggest distributing GPU-architecture specific binaries without PTX portion, instead of distributing platform-independent binaries because the former makes it difficult to perform reverse engineering. They also highlight the default parameters of NVCC compiler which generate binaries that reveal information about function declaration and parameters. Also, binaries can reveal more information than debugging tools. They further suggest strategies for thwarting binary analysis, e.g., obfuscating the function/variable names.

\subsection{Leakage in browsing and graphics applications}\label{sec:leakagewebbrowser} 

\begin{figure} [htbp]
\centering
\includegraphics[scale=0.35]{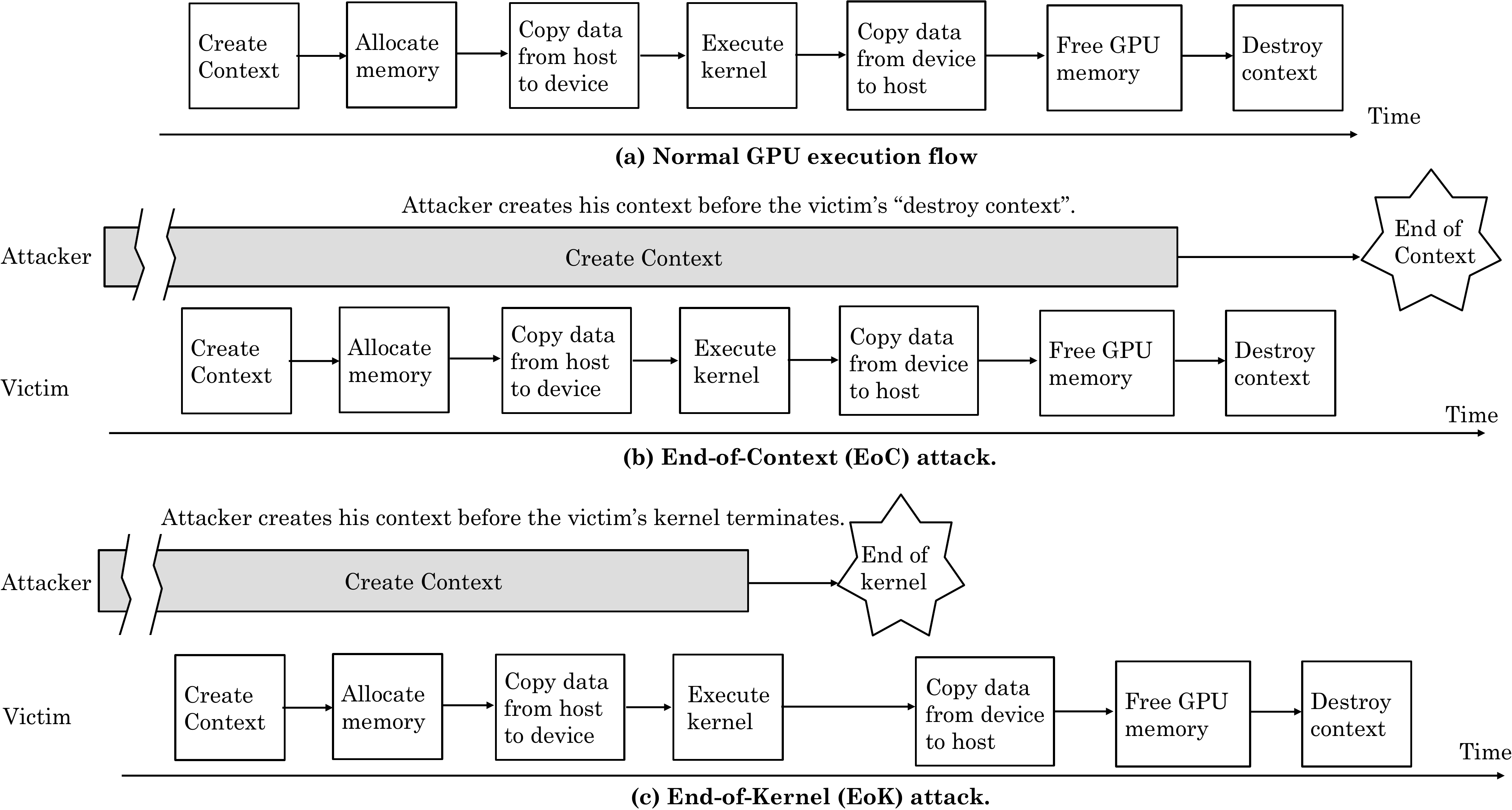}
\caption{Illustration of (a) normal execution (b) EoC attack and (c) EoK attack \cite{lee2014stealing}}\label{fig:stealingWebpagesAttack}
\end{figure}

Lee et al. \cite{lee2014stealing} note that GPUs do not erase global, shared and local memories, which makes them vulnerable to attacks. They discuss two attack points: end of GPU context (EoC) and end of GPU kernel (EoK), which attack GPU during program execution and after program termination, respectively. Figure \ref{fig:stealingWebpagesAttack}(a) shows the normal GPU execution. In EoC attack, the GPU memory is dumped after the victim frees its memory, as shown in Figure  \ref{fig:stealingWebpagesAttack}(b). Thus, if the victim does not clear its global memory, the adversary can leak global memory data, along with constant data, kernel code,  and their call-by-value arguments. Based on the memory usage history, the adversary can observe when the victim has (de)allocated global memory. 

In EoK attack, shown in Figure \ref{fig:stealingWebpagesAttack}(c), the registers (termed as ``private memory'' in OpenCL \cite{opencl_nvidiagpu})  and shared memory of victim's kernel are stolen after termination of the victim's kernel.  Since GPUs do not allow pre-emptive scheduling, long GPU programs are run using multiple same/different kernels. These kernels produce and consume intermediate results which can be leaked by EoK. In NVIDIA GPUs where the L1 cache and shared memory can exchange space \cite{harris2013sharedmemory}, EoK attack can also obtain the data from the L1 data cache. 

In EoC attack, the adversary waits till the victim program's GPU context ends and gets the final data from the global memory, if it has not been cleared by the victim program. The adversary dumps the data to the host memory only after the GPU context ends.
In EoK attack, the adversary targets to get the intermediate data between kernels of the victim program. The adversary runs a loop to dump data to the host memory after each kernel ends. The loop ends when the GPU context ends.
When multiple users are running their kernels on the GPU, the data of only the last kernel can be attacked by EoK, however, the EoC attack can succeed even when multiple victims run kernels sequentially. On AMD GPUs, the available memory size cannot be measured at runtime, and hence, they use the change in kernel execution time for detecting when a victim is using the GPU. The EoC attack is not required in AMD GPUs since an adversary can obtain the whole global memory of an AMD GPU due to its memory management features. 

They demonstrate their attacks using Firefox and Chromium web-browsers. These browsers use GPUs for accelerating webpage rendering and leave rendered webpage textures in the global memory. The textures are stored in a rearranged form using proprietary schemes. Also, due to the use of virtual and page-based memory and presence of non-color data, reconstructing the original texture is infeasible. Hence, they instead seek to infer the visited webpages based on the rearranged textures. The adversary collects memory dumps of certain webpages on the \textit{same GPU} as the victim. Then, he compares them with memory dumps of an unknown webpage. This attack can accurately guess most of the webpages. Further, if the adversary collects image snapshots of certain webpages on \textit{any GPU}, and compares them with memory dumps of unknown webpages, he can still guess nearly half of the webpages. The limitation of their technique is that it uses only side-channel data and cannot obtain raw images from GPU memories. Also, it can match the target against the websites in  a pre-decided list only. 

Zhou et al. \cite{zhou2017vulnerable} discuss an attack on GPU where a nonprivileged adversary can leak sensitive information from the remanent raw data in GPU memory. They first detect data partitions which have a high likelihood of being parts of images. The challenge in this is that the metadata of images are generally stripped. Hence, the accurate metadata such as image dimensions cannot be easily measured. Second, GPUs also store and process non-image data which makes it difficult to distinguish the image data. To resolve these issues, they use many features of image data. Initially, the adversary overwrites entire memory that it can access with a fixed value, e.g., 0xFF. Then, the malicious program executes in the background and monitors the free GPU memory space. A sudden increase in available GPU memory indicates that a victim has released GPU memory portion. The malicious program copies this and since GPU does not erase the memory before reallocation, the adversary can analyze the memory.  

Then, they split the memory dumps into fixed-size partitions and merge those partitions which are contiguous in memory space and are expected to be pixels. The partition size needs to be smaller than a single image but large enough that any large white region does not break-down the image. For example, they find that a partition size of 4K balances these tradeoffs. The partitions with all 0xFF are removed since they are unlikely to have been modified after initialization. Also, the partitions with all 0x00 are removed since these partitions have been zeroed by the OS/application. The remaining contiguous blocks are merged to form a tile. Then, remanent partitions of the victim program are extracted. Although skipping partitions with all 0xFF may leave out white spaces of an image, the remaining image is still meaningful and this justifies the choice of 0xFF as the canary. 
     
To identify partitions storing graphical data, they note that pixels are stored as four 8b channels which show red, green, blue and alpha components, respectively. The alpha component, which is 0x00 or 0xFF, shows pixel transparency. Thus, by detecting 0x00 or 0xFF, one can ascertain an opaque pixel or an unused channel. Thus, by examining the alpha value, a graphical partition can be detected. Finally, heading/trailing elements having 0x00/0xFF values are removed and middle values are retained. Generally, a tile has only single image, however, some tiles may have portions from multiple images if they are at nearby locations in the memory.  
 
Then, they detect image area in the tiles with imprecise boundaries, based on insights that (1) successive rows/columns of an image show high similarities which can be detected in the frequency domain (2) before loading into GPU, the image is decompressed and its similarities are preserved. Finally, they rearrange recovered images in proper order. They show that their technique can recover images even under high noise and after performing image transformations such as changing the contrast/brightness. They demonstrate  the recovery of the opened tabs, address bar and page-body from Google chrome, figure-portions and text-lines from recently-opened Adobe Reader documents, whole or portions of images from MATLAB.

Similarly, Zhang et al. \cite{zhang2015forensically} study the relationship between a graphic and its organization in GPU memory. Based on this, they present a scheme for visually recovering a graphic from the data stored in GPU memory.

\subsection{Leakage in virtualization and cloud-computing scenarios}\label{sec:leakagevirtualized} 

Maurice et al.  \cite{maurice2014confidentiality} study data leakage in native, virtualized and cloud computing scenarios. The procedure for testing data leakage is as follows. They use a \textit{stain} program which writes pre-determined strings in global memory and two detect programs which search the string written by the \textit{stain} program in GlM. The first detect program uses CUDA API function calls and does not require root privileges. The second detect function uses ``PCI configuration space'' which requires root privileges. Note that PCI configuration space is the medium through which the PCI can automatically configure the devices attached to its bus.

Between stain and detect functions, other actions are also performed. If the detect program can find the string, data leakage is assumed to have occurred. Table \ref{tab:dataleakScenarios} shows scenarios where leakage occurs. In native environment (see `S. No.' 1 in Table \ref{tab:dataleakScenarios}), leakage happens on user-switch, assuming persistence was ON, i.e., the driver remains loaded even when no application is accessing GPU. On soft reboot and GPU reset, leakage happens only if ECC was disabled. In other words, data erase happens as a side-effect of ECC and not due to a security scheme. No leakage happens on hard reset. Thus, GPU maintains data as long as power is turned on.

\begin{table}[htbp]
  \centering
  \caption{Overview of the attacks and results \cite{maurice2014confidentiality}. \cmark \,indicates a leak, and \xmark \, indicates no successful leak. N/A means that the attack is not applicable. $\dagger$: in cloud setup, there is not guarantee that after releasing one VM, the next VM will run on the same physical machine. $\star$: access through PCI configuration space requires root privilege. } 
     \begin{tabular}{|c|l|c|c|c|c|c|c|}\hline
        \multirow{2}[0]{*}{S. No.} & \multirow{2}[0]{*}{Setup} &       \multirow{2}[0]{*}{ECC}  & \multicolumn{5}{c|}{Actions between taint and search } \\ \cline{4-8}
      & &   & switch user     & soft reboot     & reset GPU     & kill VM and start another     & hard reboot\\ \hline
    \multicolumn{8}{|c|}{Access using CUDA API functions} \\ \hline
     \multirow{2}[0]{*}{1} & \multirow{2}[0]{*}{Native} & on    &\cmark     & \xmark     & \xmark   & \multirow{2}[0]{*}{N/A} & \xmark \\ \cline{3-6} \cline{8-8}
       &   & off   & \cmark     & \cmark     & \cmark     &       & \xmark \\  \hline 
    \multirow{2}[0]{*}{2}& \multirow{2}[0]{*}{Virtualized} & on    & \cmark     & \xmark     & \xmark     & \xmark     & \xmark \\ \cline{3-8}
    &      &  off  & \cmark     & \cmark     & \cmark     & \cmark     & \xmark \\ \hline
    \multirow{2}[0]{*}{3}& \multirow{2}[0]{*}{Cloud} & on    & \cmark     & \xmark     & \xmark     & \multirow{2}[0]{*}{N/A$\dagger$} & \multirow{2}[0]{*}{N/A} \\ \cline{3-6} 
    &       & off   & \cmark     & \cmark     &  \cmark    &       &  \\ \hline
    & \multicolumn{7}{|c|}{PCI configuration space access} \\ \hline
     \multirow{2}[0]{*}{4}& \multirow{2}[0]{*}{Native} & on   & \multirow{2}[0]{*}{N/A$\star$} & \xmark     & \xmark     & \multirow{2}[0]{*}{N/A} & \xmark \\ \cline{3-3}\cline{5-6}\cline{8-8}
     &       & off   &       & \cmark     & \cmark     &       & \xmark \\ \hline
     5 & Virtualized & -     & N/A$\star$  & \xmark     & \xmark     & \xmark     & \xmark \\ \hline
    6 & Cloud & -     &N/A$\star$   & \xmark     & \xmark     & N/A$\dagger$  & N/A \\ \hline
     \end{tabular}%
  \label{tab:dataleakScenarios}%
\end{table}%

In a virtualized environment (see `S. No.' 2 in Table \ref{tab:dataleakScenarios}), the attacker has full control on the VM and its virtualized GPU. Here, the results are similar as that in a native environment. To study the impact of the hypervisor, they generate a guest VM, run stain function and then destroy the VM. Then, they generate another VM and run detect function. They note that the second guest VM can read the data left by the first guest VM. Clearly, the hypervisor fails to ensure isolation between VMs. The results in a cloud environment (see `S. No.' 3 in Table \ref{tab:dataleakScenarios}) are similar to that in a virtualized environment. Thus, in the cloud, if the attacker can run the VM on the same machine as the previous user, he can leak the data of the user.

Accessing GPU through CUDA API provides only a partial view of the memory which is accessible by MMU. An attacker with root-privilege can use PCI configuration space to access the entire memory, bypassing the MMU. In native environment (see `S. No.' 4 in Table \ref{tab:dataleakScenarios}), the leakage was observed after soft reboot and GPU reset, but not after a hard reset. However, no leakage was observed in virtualized and cloud environment (see `S. No.' 5 and 6 in Table \ref{tab:dataleakScenarios}). 

They note that erasing the memory at allocation time avoids leakage through CUDA APIs, but not through PCI, because in this case, no memory allocation is performed. Hence, it is better to erase the memory at the time of deallocation. Also, cloud providers can erase memory before allocating an instance to the user. Finally, a user can erase the memory before freeing it, although this incurs a performance penalty and may not erase entire memory due to issues such as fragmentation and indeterminism in the behavior of CUDA memory manager.

\subsection{Detecting leakage through taint-analysis}\label{sec:taintanalysis}
Taint-analysis tracks the flow of ``tainted'' data, i.e., sensitive input that transforms over the course of execution to affect memory locations. This allows detecting the  information-leakage, clearing the sensitive data after their scope and also identifying the presence of malware. 

Hayes et al. \cite{hayes2017gpu} implement a GPU taint-tracking scheme which uses static binary instrumentation for performing dynamic taint tracking of GPU applications. They divide instruction operands into two types, viz., taintable and untaintable, based on whether  they can be tainted at runtime.  Examples of untaintable operands include built-in variables such as thread ID, block ID, block dimension, pointer-type kernel parameters, user-specified constants and loop-induction variables. Examples of taintable sources are image, plain-text, encryption key, etc.  Every thread tracks information-flow on its own. 

Their analysis proceeds iteratively and each iteration has two steps. In forward step, taintable/un-taintable operands are flagged at the points of taint state-transition, e.g., storing the constant value in a register switches it to the  untaintable type. In backward step, operands which can affect memory value are flagged. Remaining variables cannot flow to memory even if they are taintable, such as stack-frame pointers, loop-counters, etc. Then, only those operands which are flagged in both forward and backward steps are tracked, and this reduces the overhead of taint-tracking significantly.  

The taint record is stored in GPU registers, which reduces its access latency compared to storing it in LoM or GlM. Further, compared to CPUs, GPUs have much larger register file  \cite{mittal2016SurveyCPURF,mittal2016SurveyGPURF} and not all registers are used by most kernels. Since registers/ShM/LoM/GlM have different lifetimes, their scheme handles them differently and clears them at different times. LoM and GlM  taints are cleared at the completion of each kernel and program, respectively. The ShM is cleared after all threads of a block complete their work. Registers are cleared by each thread before exiting.

Their scheme does not require VM emulation or a dynamic instrumentation framework which is generally unavailable on GPUs. Their approach enables  protection against the register-spilling based attack \cite{pietro2016cuda} and incurs only small performance overhead. Further, their approach allows tracking data transfers between CPU and GPU, which is especially useful in fused CPU-GPU systems. Furthermore, by using their approach, a suspicious program can be prevented from accessing uninitialized data left by other programs/users. 

\subsection{Protecting sensitive data using system management mode}\label{sec:kimSMM}

Kim et al. \cite{kim2016demand} present a booting scheme which uses SMM for isolating the GPU kernel and key of cryptographic algorithms.
Some works store the key in GPU registers \cite{vasiliadis2014pixelvault}.  
However, since registers are not shared between threads, every thread needs to reserve multiple (32b) registers to store the key, which degrades performance by limiting the maximum number of  kernels that can run concurrently. By comparison, storing the key in GPU cache is better since it cannot be accessed by any CPU application and its contents cannot be accessed after completion of GPU kernel. However, since the cache is hardware-managed, it is important to ensure that the key does not get evicted from the cache. To avoid this issue, they store the key in the constant cache before booting.

However, by compromising the OS kernel, the adversary can launch code-injection attack to tamper GPU kernel code. To subvert this, they use SMM for isolating authenticated GPU kernel in instruction-cache. Thus, only SMI and trusted GPU kernel can access the key. Before the end of SMM, copies of kernel and key in GPU memory are cleared. After program completion, the results can be copied from GPU memory, but kernel-code and key are invalidated.  
The sensitive data is handled only in SMM, and thus, security is ensured.  
  
They note that the keys are always stored at a fixed address in GPU memory and based on this, the exact address for storing the keys in constant memory can be obtained.  The size of L1/L2/L3 constant caches are 1KB/8KB/32KB, respectively and the L2/L3 caches are shared with instruction memory. Hence, if the total key size exceeds L1 constant cache capacity, it may get evicted.  However, they note that keys are evicted only when total key size exceeds 47,520 bytes and thus up to 270 AES keys of 176 bytes each can be accommodated easily. During booting, security-unrelated and security-related tasks are performed in regular CPU mode and SMM, respectively. This reduces the overhead of SMI program and also allows leveraging the functionality of GPU driver and OS kernel which are suspended in SMM.  Their technique protects the key even when the kernel is compromised and its performance penalty is negligible. The limitation of their technique is that an adversary with physical access to the processor can subvert secure memory using a cold-boot attack.

\section{Side- and Covert-channel attacks}\label{sec:sidechannel}
\subsection{Timing side-channel attack}\label{sec:timingchannelattack}

In this section, we discuss works that demonstrate timing side-channel attacks   along with their countermeasures. Table \ref{tab:timingChannelAttack} highlights their similarities and differences. 

\begin{table}[htbp]
  \centering
  \caption{Comparison between techniques of Jiang et al. \cite{jiang2016complete,jiang2017novel} and Kadam et al. \cite{kadam2018rcoal}}
 
    \begin{tabular}{|p{2.7cm}|p{6cm}|p{7cm}|}
    \hline
          & Jiang et al. \cite{jiang2017novel} & Jiang et al. \cite{jiang2016complete} and Kadam et al. \cite{kadam2018rcoal} \\
    \hline
    Encryption algorithm & \multicolumn{2}{l|}{128-bit ``electronic codebook'' mode AES encryption with T-tables. It uses 16B key to encrypt a 16B block}    \\
    \hline
    Table stored in  & Shared memory &  Global memory  \\
    \hline
    Correlation & Execution time depends on shared memory conflicts, which depends on the address accessed by each thread & Execution time depends on the number of unique cache line requests after coalescing, which depends on the address accessed by each thread  \\
    \hline
    \end{tabular}%
  \label{tab:timingChannelAttack}%
\end{table}%

\begin{figure} [htbp]
\centering
\includegraphics[scale=0.40]{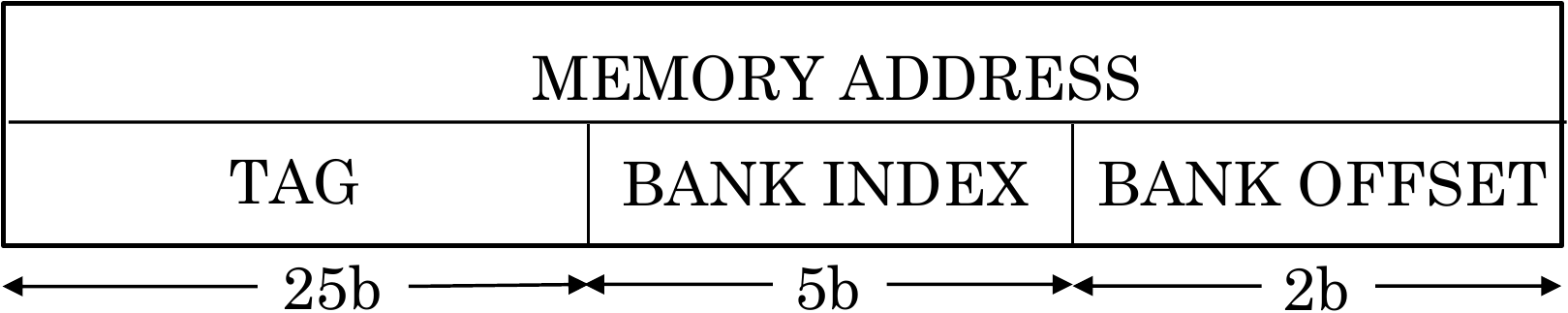}
\caption{Mapping of memory address to ShM bank \cite{jiang2017novel}  }\label{fig:MemoryAddressDivision_ShM}
\end{figure}

Jiang et al. \cite{jiang2017novel} present a differential timing attack which leverages timing variability due to ShM bank conflicts to recover AES encryption key. The AES-128 algorithm includes ten rounds of operations and in each round, a 16B key is used, which is termed as the `round-key'. From any round key, an adversary can obtain the original 16B key. Their attack focuses on leaking the last-round key. Figure \ref{fig:MemoryAddressDivision_ShM} shows the address division for mapping to ShM. Accesses going to the same bank cause conflict and due to the SIMT execution model, the bank seeing the highest number of conflicts determines the execution time of the memory instruction. Hence, the execution time of an ShM instruction depends on the memory addresses and consequent bank-conflicts.

They use a kernel which loads data from ShM using a thread-warp. The kernel is written in a manner to create conflicts among the threads. The address accessed by each thread decides the number of conflicts and hence, the completion time of the load operation. In the AES algorithm, memory access for reading the lookup table depends on the value of the key. Since the table is word-aligned, the completion time of a table access for a warp is expected to vary linearly with the number of conflicts generated by the threads. The overall encryption time also depends on the conflicts caused by table accesses. 
  
Due to the relation between table-index and round-key, the number of conflicts in one round of AES operation for a warp can be predicted by guessing the round-key. If an adversary uses many plaintexts and correlates the encryption latency with the number of conflicts, then, this correlation should be high for a correct key guess and low for wrong key guesses.  This is the basis of launching differential timing attack. Since each table access uses one state byte, each round key byte can be separately attacked. For each guess of the byte value (between 0 to 255), they compute the correlation between average timing and number of bank conflicts. Based on this, a correct key byte is distinguished from incorrect ones. In practice, they compute average timing for data samples leading to two and four conflicts. Then, the ``difference of means'' between them is expected to be twice the penalty for one conflict, which is 19 (\ApproxSign 2*9.8) cycles, where 9.8 cycles is average penalty for one conflict. Proceeding in this way, they recover all the 16 bytes of the key.  Their attack is also applicable to other table-based cryptographic algorithms. To mitigate their attack, use of ShM can be completely avoided, or the use of ShM data can be reduced to avoid all bank conflicts. 
 
They also note that encryption time measurements obtained on GPU are more accurate than that on CPU due to the data-transfer and initialization operations. Further, they test the scalability of their attack by attacking an 8192-thread AES implementation. In such a real scenario, the timing has to be measured on CPU since the kernel code running on GPU cannot be modified to insert timing routines. Also, since the scheduling of blocks/warps on GPU SMs is not known, launching an attack to recover accurate key becomes challenging.

Jiang et al. \cite{jiang2016complete} show a correlation timing attack on GPU to completely recover the AES key. They use 128-bit ``electronic codebook'' (ECB) mode AES encryption with T-tables, which uses 16B key to encrypt a 16B block. Each thread performs one block-encryption. They store the keys in the constant cache. All threads of a warp simultaneously read the same round-key. T-tables are stored in GlM since access to them leads to different memory requests which are serialized and hence, storing them in the constant cache would lead to resource wastage. They record time in two ways: clean measurement, where the warp execution time inside a kernel can be measured, and noisy measurement, where the latency of incoming/outgoing messages can be measured. 

They note that the AES algorithm issues memory requests for loading its T-table entries, whose addresses depend on the plaintext and encryption key. Under SIMT paradigm, a load instruction of a warp generates one memory request from each of its 32 threads. These requests are coalesced and merged with the requests queued in miss status holding register (MSHR). The time incurred in serving the 32 memory address requests of a warp scales linearly with the number of unique cache line requests. Given this high correlation, memory addresses (and hence, the key) can be inferred from the timing measurements. Specifically, in the last round of AES, each table index can be obtained from one byte of the key and corresponding byte of ciphertext, irrespective of other bytes of ciphertext. Using this, each byte can be individually guessed. For each key guess, they compute number of coalesced accesses (NCAs) for every 32-block message, and then, find the correlation of timing with the NCAs.  For a correct key-byte guess, the NCA is correct and hence, the correlation is highest, otherwise it is low. From this, the correct key can be estimated.

They note that with an unoptimized binary, the correlation is even higher. With no optimizations, the CUDA instructions are not reordered. Hence, a table access gets stalled due to data dependency, which makes timing measurements more predictable. Their attack can recover all key bytes of AES-128 in 30 minutes. Under noisy measurements, only 10 out of 16 key bytes are recovered. However, by increasing the number of measurements and using noise-reduction strategies, all the 16 key bytes are recovered and thus, the addition of noise does not fully mitigate the attack. 

To mitigate these attacks, more noise/randomness can be added to timing measurements and the key can be frequently changed. Also, the mapping of the table-lookup index to cache line can be randomized, which prohibits the attacker from deducing the number of unique cache line requests generated.

Kadam et al. \cite{kadam2018rcoal} note that the determinism in coalescing mechanism leads to a security vulnerability. They assume the same attack model as Jiang et al. \cite{jiang2016complete}. They note that the execution time of last-round and overall execution time have a high correlation with last round coalesced accesses. Based on this, the correct key can be found as one showing the highest correlation. They assume that the attacker can observe last-round execution time and thus, launch a stronger attack. Also, use of last round execution time  allows them to evaluate their technique on the simulator and experiment with different coalescing policies.

Correlation timing attack exploits the deterministic nature of memory access coalescing, from which the number of coalesced accesses (NCA) can be accurately inferred. Disabling coalescing thwarts this attack, but exacerbates data-migration and energy overheads. GPU AES implementation leaks information due to two reasons. Firstly, coalescing groups all threads of a warp in a single subwarp which makes it easy to infer NCA by (1) finding the table indices and (2) finding the number of memory blocks (i.e., NCA), based on the sequential mapping of table entries to memory blocks of known size.
Secondly, the grouping-order of the thread does not influence coalescing since all threads of a warp are considered together for coalescing. On performing coalescing at the subwarp level, the thread-grouping order affects NCA, depending on the threads falling into the same subwarp.

\begin{figure} [htbp]
\centering
\includegraphics[scale=0.35]{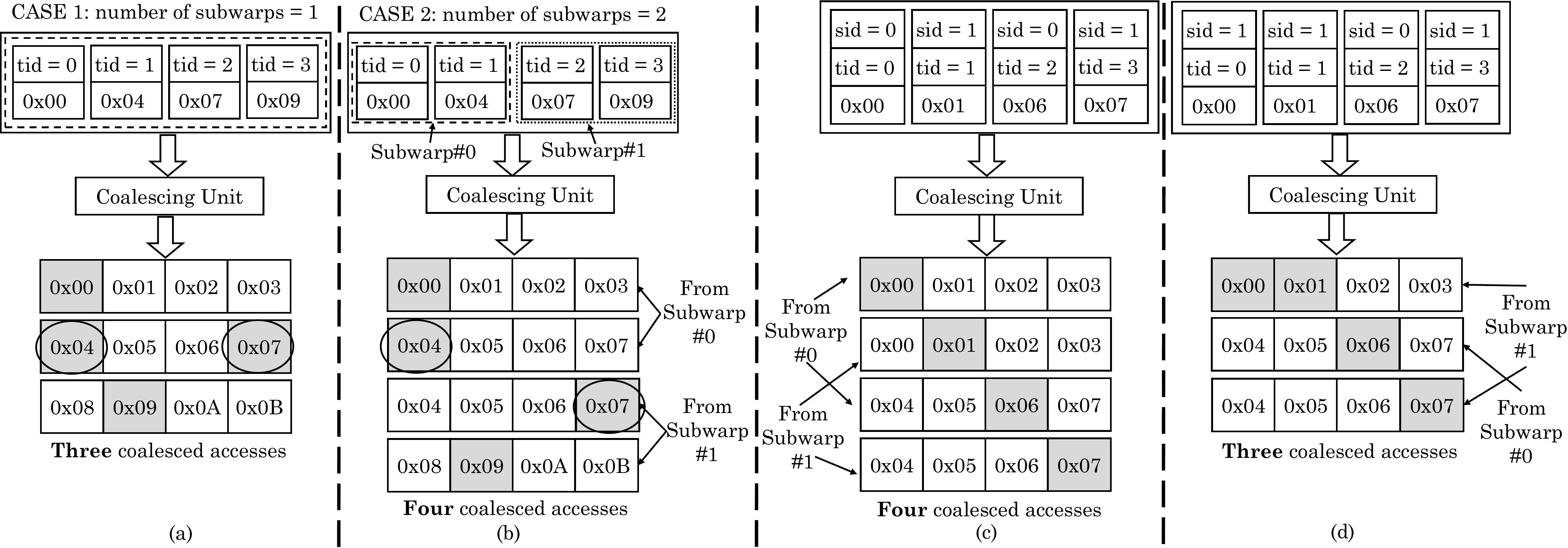}
\caption{ (a)-(b) Effect of subwarps on  coalescing \cite{kadam2018rcoal}. (c)-(d) Effect of different security schemes on coalescing with 2 subwarps. (c) FSS+RTS and (d) RSS+RTS. {\tt sid} and {\tt tid} show subwarp ID and thread ID, respectively }\label{fig:RCoal_TimingChannel}
\end{figure}

Figures \ref{fig:RCoal_TimingChannel}(a)-(b) illustrate the impact of subwarps on memory coalescing. For simplicity, assume that a warp has four threads, which may be grouped in a single subwarp (Figure \ref{fig:RCoal_TimingChannel}(a)) or two subwarps (Figure \ref{fig:RCoal_TimingChannel}(b)). Each thread generates one memory access request. With a single subwarp, the requests from the second and third thread are coalesced and thus, a total of three coalesced accesses are generated. On using two subwarps, coalescing is performed separately for each subwarp. Each subwarp generates two accesses, for a total of four accesses.  

Their techniques tackle these limitations to reduce the correlation and make it difficult to infer the last round key bytes.  The first technique, termed ``fixed-size subwarps'' (FSS) changes the number of subwarps since without knowing this, an adversary may not accurately estimate NCA. With increasing number of subwarps, the variance in NCA reduces, which also weakens the correlation. The limitation of this technique is that, since different values of subwarp-count lead to significantly different execution time, an adversary can guess the subwarp-count by repeatedly measuring the execution times.

The second technique, termed ``random-sized subwarp'' (RSS),  randomizes the number of threads per subwarp while keeping the total number of threads per warp still constant, i.e., 32. This makes it difficult for an adversary to guess NCA even if he knows the number of subwarps. The third technique, named ``random-threaded subwarp'' (RTS) randomizes the threads which form a subwarp and thus, brings further randomness in NCA. The third technique can be combined with previous two techniques.

Figures \ref{fig:RCoal_TimingChannel}(c) and \ref{fig:RCoal_TimingChannel}(d) show examples of FSS+RTS and RSS+RTS, respectively. With FSS+RTS, both subwarps have  2 threads, but they are not mapped in order, e.g., subwarp 1 has threads 1 and 3, and not threads 2 and 3. Hence, four coalesced accesses are produced. With RSS+RTS, first and second subwarps have 1 and 3 threads, respectively.   This changes the mapping of one thread, viz., thread 0 is now mapped to subwarp 1. Hence, three coalesced accesses are generated. Clearly, RSS+RTS provides randomness for stronger security. For the example shown in Figure \ref{fig:RCoal_TimingChannel}(d), RSS+RTS 
also provides the benefit of reducing NCA, however, in general, it may not always reduce NCA. Their techniques improve GPU security significantly and allow achieving a tradeoff between security and performance loss.

\subsection{Power side-channel attack}
    
Luo et al. \cite{luo2015side} present a side-channel attack on GPU for extracting the last round-key of AES running on a GPU. They note that SIMT execution of GPU leads to high amount of noise in a power side-channel. This is because undocumented scheduling policies introduce randomness in the execution order of threads/warps and hence, at any point in time, different threads can be in different execution phases. Further, a GPU has multiple SMs, which simultaneously perform independent encryptions. Hence, it is difficult for an attacker to get synchronized and clean power traces. 
 
\begin{figure} [htbp]
\centering
\includegraphics[scale=0.35]{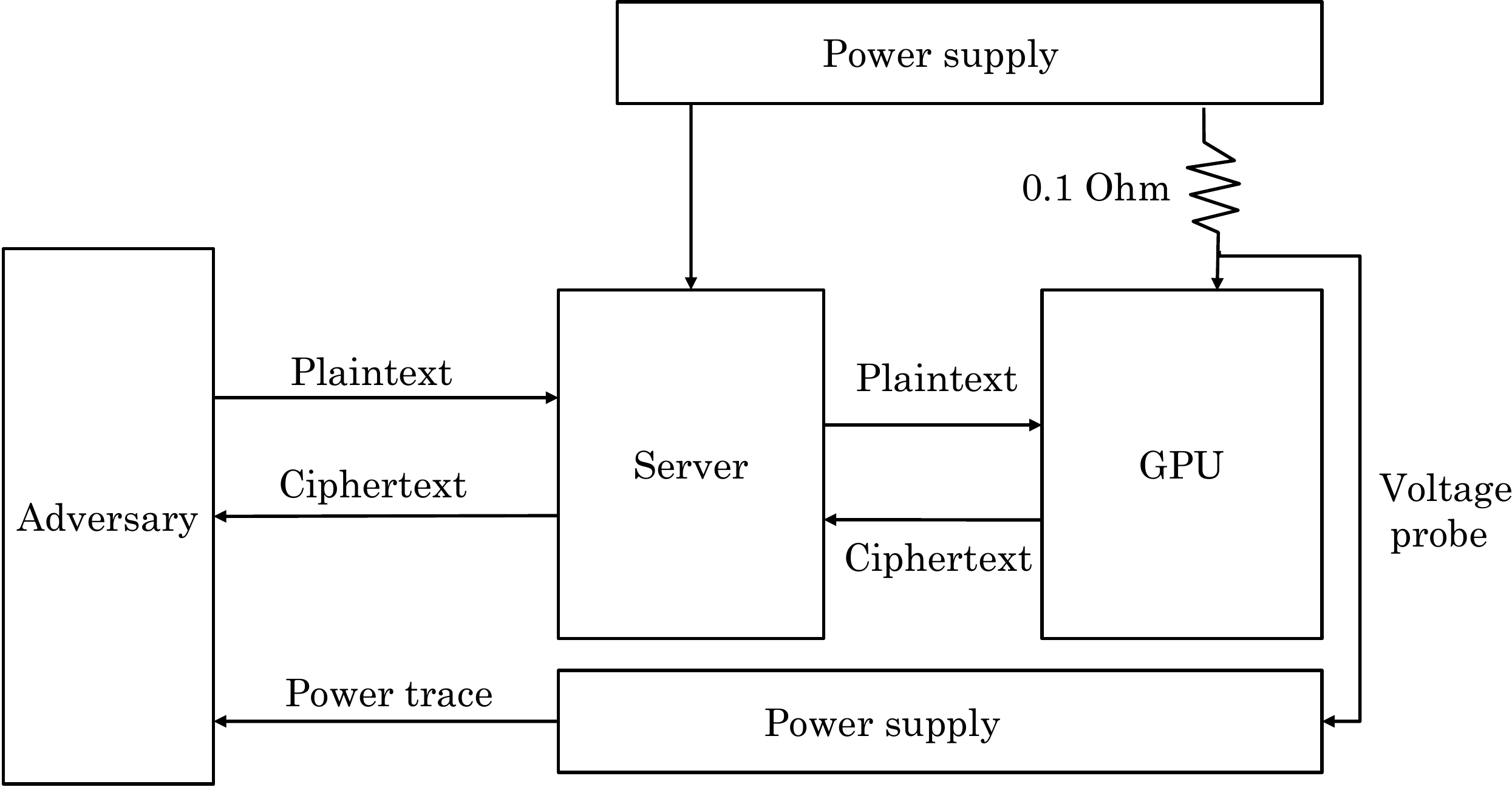}
\caption{ The power measurement setup used by Luo et al. \cite{luo2015side} }\label{fig:PowerSideChannel}
\end{figure}

To address these challenges, they first develop a power model to estimate GPU power consumption as a function of the round-key. They use it to generate power traces during encryption for different inputs. The experimental setup for measuring power is shown in Figure \ref{fig:PowerSideChannel}. They also remove the sources of noises, such as cooling fan, and increase the number of traces to improve the signal-to-noise ratio. Using their strategy, they obtain clean power traces. Then, for each byte of the key, they make 256 guesses and obtain their correlation with the power consumption. The correlation value was found to be high for a correct guess. This confirms that  CUDA AES running on a GPU is vulnerable to correlational power attack. They also note that in any encryption, if the variation in data of different blocks of plaintext is high, the number of traces required for an attack to succeed increases, and thus, the attack becomes more difficult.

\subsection{Covert channel attack}\label{sec:covertchannel}

Naghibijouybari et al. \cite{naghibijouybari2017constructing} note that, as GPUs begin to support multiprogramming, they also become vulnerable to covert-channel attacks   between two concurrently running kernels. 
They assume that the trojan and the spy kernels, which belong to two different applications,  mutually communicate covertly. To ensure  that these kernels co-reside and share resources, they first reverse engineer the scheduling policies of the block and warp schedulers. They note that if the number of blocks in every kernel equals or exceeds the number of SMs on GPU such that no block exhausts the resources of an SM, they can concurrently-reside in an SM.  
The contention on SM is also determined by another level of sharing. Every warp is associated with one of the multiple warp schedulers present on every SM. Warps sharing the same scheduler have different degrees of contention due to  the competition for the issue BW assigned to the scheduler. Since warps are assigned to the warp schedulers in a round-robin manner, the kernel parameters of trojan and spy can be chosen to co-locate them on the same SM and even the same warp scheduler. 
   
They discuss three types of CCs: through caches, functional units and GlM. Cache CC is demonstrated using constant cache, since the small size of L1 and L2 caches allows easy creation of contention. They first deduce the characteristics of constant memory and cache at different levels of hierarchy. For CC through L1 cache, they run two concurrent kernels using two streams on GPU. The trojan kernel encodes `1' or `0' by causing contention or staying idle, respectively. To cause contention on one set, the trojan accesses a single set.  The spy also accesses the same set and records the latency. A high latency implies that a value `1' was transmitted since the data was replaced by a trojan and low latency implies transmission of value `0'. 
  
When co-residency of two kernels on same SM cannot be realized, communication can be achieved through L2 constant cache, which is shared between all SMs. CC creation on the L2 cache is similar to that on L1 cache, except that the size of array accessed by spy/trojan are different due to different cache parameters.  
     
For creating CC through FUs, they observe that contention occurs only between warps belonging to the same warp scheduler. Also, to create contention, both kernel and spy need to send instructions to the same special functional unit (such as the FU performing {\tt \_\_sinf} operation). They ensure that the thread-blocks of the two kernels are co-located on each SM. The spy kernel executes multiple {\tt \_\_sinf} operations. To send `0', the trojan stays idle, and hence, spy observes a latency corresponding to  no contention. To send `1', the trojan executes  {\tt \_\_sinf} operations, and hence, its warps are scheduled to issue instructions along with the spy warps. In this case, the spy observes higher latency indicating presence of contention. Extending the above observation, kernels on different warp schedulers can create parallel CCs for increasing BW.

Creating CC through GlM is suitable to provide extra contention when kernels do not reside on the same SM. Due to the high GlM  BW, use of normal load/store instructions does not create contention. Hence, they use atomic operations which cause contention due to a limited number of atomic units. The trojan and spy access two different arrays in GlM and perform atomic operations on a specific GlM address. Although atomic operations are slow, CC through them achieves BW which is comparable to that with other CCs. Table \ref{tab:covertchannelsummary} summarizes the characteristics of CCs through different processor-components and BW achieved by each CC.

\begin{table}[htbp]
  \centering
  \caption{Conditions for constructing CC through various components and the  BW achieved (a ballpark figure) \cite{naghibijouybari2017constructing}. Two kernels refer to trojan and spy kernels.}
    \begin{tabular}{|l|p{12cm}|c|}
    \hline
          & \multicolumn{1}{|c|}{Condition} & BW achieved \\
    \hline
    L1 Cache & Both kernels should reside on the same SM & 40Kbps \\
    \hline
    L2 cache & Both kernels use shared L2 cache but may not not reside on same SM & 20Kbps \\
    \hline
    FU    & Thread-blocks of two kernels should co-reside on every SM since contention happens only between warps of the same warp-scheduler. Also, both kernels need to issue an operation to the same special functional unit & 30Kbps \\
    \hline
    GlM   & Both kernels perform atomic operations on two different arrays located in global memory & 40Kbps \\
    \hline
    \end{tabular}%
  \label{tab:covertchannelsummary}%
\end{table}%

They further discuss strategies for increasing CC BW by achieving synchronization. This avoids loss of BW due to drift between kernels or sequential launch of kernels. Also, they parallelize the transfer to allow communication between multiple trojans and spies. Further, to remove noise in CC due to co-location of other workloads, they modify block-scheduler such that only  trojan and spy kernels and no other workload can run on the same SM. For example, if the spy requests maximum amount of ShM (or threads, registers, etc.)  and the trojan does not request ShM, then they can be co-located on the same SM, but no other workload requiring ShM can co-locate on the same SM. 

To thwart these CCs, GPU resources can be partitioned so that communicating kernels cannot measure mutual contention. For example, cache partitioning  \cite{mittal2017SurveyCachePart} can be used or co-location of different kernels can be avoided. While incurring performance overhead, these approaches can thwart side and covert channel attacks due to uncontrolled access to shared resources. Similarly, by allowing pre-emption, introducing randomness in scheduling policies and noise in latency-measurements, the challenges in launching the attack can be increased. Also, cache bypassing \cite{mittal2016SurveyCacheBypassing} can be used to prevent sensitive data from being stored in the cache.

\section{Malware, Buffer-overflow and Denial-of-service Attacks}\label{sec:malwareOverflowDoS}
In this section, we discuss active attacks, viz. use of GPU malware (Section \ref{sec:malware}),  causing buffer-overflow (Section \ref{sec:bufferoverflow}) and denial-of-service (Section \ref{sec:denialofservice}).  

\subsection{Malware attack}\label{sec:malware}

Zhu et al. \cite{zhu2017understanding} demonstrate that discrete GPUs cannot be trusted as secure co-processors, rather, they may host stealthy malware. They study the limitations of PixelVault \cite{vasiliadis2014pixelvault} technique   which assumes  GPU hardware registers to be secure storage. PixelVault presents a GPU-accelerator for AES and RSA encryption. PixelVault assumes that after the kernel is launched, the attacker can control entire system and run the program with any privilege. Their work is based on assumptions shown in column 2 of Table \ref{tab:pvassumptionrefute} since these assumptions ensure certain properties (column 3). However, Zhu et al. refute these assumptions (column 4).

\begin{table}[htbp]
  \centering
  \caption{PixelVault's \cite{vasiliadis2014pixelvault} assumptions and their security benefits and why they don't hold \cite{zhu2017understanding}}
    \begin{tabular}{|l|p{4.5cm}|p{5.0cm}|p{5.5cm}|}
    \hline
    No. & Assumptions & Benefits from assumption & Why assumption does not hold \\
    \hline
    1 & If the code of a running kernel is stored entirely in I-cache, it cannot be replaced. & It ensures that an attacker cannot substitute PixelVault GPU code without stopping it and hence, without losing the master key stored in GPU registers.  & The MMIO registers on NVIDIA GPUs can flush I-cache and thus, replace code of running kernels with malicious code  \\
    \hline
    2 & GPU register values cannot be obtained after the kernel terminates.  & This prohibits an attacker from obtaining a master key by stopping a running kernel. &     On launching two kernels on different CUDA streams, {\tt cuda-gdb} can leak keys stored in registers of the terminated kernel as long as the other kernel is running.    
    \\
    \hline
    3 & Unless compiled with debug support, a running kernel cannot be stopped and debugged. & It ensures that attacker cannot read registers by attaching a GPU debugger to the running PixelVault kernel. & Latest CUDA runtime allows attaching a debugger to a running kernel using root privileges.  \\
    \hline
    \end{tabular}%
  \label{tab:pvassumptionrefute}%
\end{table}%
   
They show that violation of these assumptions and unrestricted DMA access from GPU to CPU memory allow subverting the protection of IOMMU. Using this, an adversary can run stealthy malware on GPU and thus, systems based on PixelVault's approach are not secure. Note that IOMMU monitors devices' DMA to system memory for protecting them from illegitimate accesses. The IOMMU prohibits a device from accessing CPU pages which are not mentioned in its I/O page table. 
 
They further show two attacks. The first attack is on the in-kernel NVIDIA GPU driver. When the driver is being loaded and used by OS kernel, they binary-patch it. The patched driver maps sensitive CPU memory into address-space of an unprivileged GPU application. The second attack combines the microcode running on a microprocessor with malicious code to access any portion of CPU memory.

Balzarotti et al. \cite{balzarotti2015impact} divide the GPU malware into three categories based on the OS privileges required by them and the amount of knowledge they possess about internal details of the GPU driver. (1) \textit{User-space malware} have only normal privileges. They perform regular computations using standard GPU commands, without relying on any software bug. (2) \textit{Super-user malware} need elevated privileges and hence, can execute additional tasks which are not possible through user space. However, the adversary has no knowledge of the GPU driver/card. (3) \textit{Kernel-level malware} not only have elevated privileges but also know the internals of the graphic driver. Hence, they can tamper the data structures in the kernel.

They further describe four strategies which the malware can adopt to avoid detection by forensic tools. (1): ``Unlimited code execution'': due to non-preemptive nature of GPU, a single task can occupy the GPU completely. To avoid this, GPU driver uses a timeout scheme (e.g.,  {\tt hangcheck} function), by which a long-lasting process can be killed. However, the malware can disable this to occupy the GPU indefinitely. (2) ``Process-less code execution'': a GPU kernel is usually always controlled by a host process. However, the malware may run a kernel without any controlling process on the host  by killing the host process right after the GPU kernel starts.

(3) ``Context-less code execution'': The ``process-less execution'' still leaves traces in the GPU driver, e.g., the buffer objects and hardware context of the GPU kernel may still be present in the driver's memory. In ``context-less'' execution, the hardware context of GPU kernel is removed completely from the records of the driver. Thus, the kernel can hide its presence completely. Both process-less and context-less executions require that the kernel has super-user privileges and has already achieved ``unlimited code execution''. However, the latter additionally requires knowledge about driver internals. (4) ``Inconsistent memory mapping'': Generally, the list of accessible physical pages is kept both in the OS and GPU memories. However, the information kept in these two-page tables can be made to differ. Then, the OS page table points to the correct page, but the GPU address points to a random memory location. 

Table \ref{tab:antiforensic} summarizes the characteristics of these strategies along with the privileges and information needed by them. 
The last three columns show the results of memory forensic, whether it can find (1) the processes which are using GPU, (2) the code executed by them and (3) the memory regions accessible to these kernels. The \cmark sign shows that the forensic goal can be realized by inspecting only system memory. If so, the analysis may need information of only the OS data-structures or driver internals, and this is indicated as (OS) or (driver), respectively. The \xmark sign shows that the corresponding   forensic analysis is not possible.  They also recommend that checking the graphic page tables, hangcheck flag, list of buffer objects, contexts and register files can provide an insight into the type of malware present.

\begin{table}[htbp]
  \centering
  \caption{Characteristics of anti-forensic strategies \cite{balzarotti2015impact}}
    \begin{tabular}{|l|l|l|l|l|}
    \hline
    \multicolumn{1}{|c|}{\multirow{2}[4]{*}{Strategy}} & \multicolumn{1}{c|}{\multirow{2}[4]{*}{Privilege/knowledge requirement}} & \multicolumn{3}{c|}{Forensic objectives of listing} \\
\cline{3-5}          &       & processes & kernels & memory regions \\
    \hline
    Unlimited execution & Superuser & \cmark (OS) & \cmark (driver) & \cmark (OS) \\
    \hline
    Process-less execution & Superuser & \xmark & \cmark (driver) & \cmark (driver) \\
    \hline
    Inconsistent map & (Superuser, driver-internals) & \cmark (OS) & \cmark (driver) & \xmark \\
    \hline
    Context-less execution & (Superuser, driver-internals) & \xmark & \xmark & \xmark \\
    \hline
    \end{tabular}%
  \label{tab:antiforensic}%
\end{table}%

Danisevskis et al. \cite{danisevskis2013dark} demonstrate an attack whereby an adversary can exploit DMA to bypass GPU memory protection schemes  and gain access to privileged regions of the memory. They demonstrate these attacks on an Android smart-phone with sound memory isolation and security mechanisms. The adversary tricks a benign user into installing a malicious application which initially requests only normal access rights. This application seeks to escalate its privilege by patching the text section of the kernel. 
 
For this, the application patches the {\tt sys\_reboot} system-call which is called only on power-down, and thus, patching of this is likely to go unnoticed. Then,  the physical address for patching is determined and   the above system-call is attached to the session address space of the mobile-GPU. Then, they configure the GPU to change the caller's user ID to zero (or root) and write this to text section of the kernel. Based on this, they prepare a rendering job and submit it to the kernel driver. Their work confirms that DMA-based malware can pose security threats for mobile GPUs.

Keyloggers are malware that stealthily log keyboard activity for stealing sensitive data. Ladakis et al. \cite{ladakis2013you} implement a keylogger malware on GPU which tracks the keyboard buffer directly from GPU through DMA without modifying kernel's code or data structures except for that page table. Their keylogger has two modules. A CPU-based module runs only once at the time of booting. It finds the address of keyboard buffer in physical memory, since this address may randomly change on every system boot. They implement a loadable kernel module which scans entire host memory to find this address, although its limitation is that the search time increases with increasing physical memory size.

Then, a GPU-based module continuously tracks changes in this keyboard buffer. For this, the GPU needs to have access to the buffer. Since NVIDIA GPUs share an address space with the host process, the buffer must be mapped in the virtual address space of the host process. To achieve this, the pages where buffer resides can be included in the page table of the host process. A keystroke monitoring interval of 100ms incurs small overhead while still allowing  the capture of   keystrokes of even fast typists. Their keylogger can track all keystrokes, store this data in  the large GPU memory and process it in real-time using GPU's massive compute-power.

\subsection{Buffer overflow attack}\label{sec:bufferoverflow}
In the absence of memory-access protection mechanisms,  an adversary can read/write memory regions to either leak or destroy data. Miele \cite{miele2016buffer} and Di et al.  \cite{di2016study} discuss strategies for launching buffer-overflow attacks,  their impact and limitations. Erb et al. \cite{erb2017dynamic} propose the use of canaries to mitigate buffer-overflow attacks.

Miele \cite{miele2016buffer} discusses ways of causing stack and heap overflows for corrupting sensitive data or changing the execution flow. They discuss two attacks. In the first attack, a function pointer in static memory is overwritten to cause undefined behavior. They consider an array with a certain starting address and a predefined maximum length. It is assumed that the address of the function to be executed is stored in the address after the last possible element in the array. If the malicious user can overwrite beyond the length of the array, causing the buffer (or array) to overflow, the address of the function to be executed can be replaced by a different value, such as the address of a malicious function. Through this, execution can be made to jump to any address in the code memory. However, they note that simple injected code execution was found to be infeasible as the code and data address spaces are separated. Similarly, overwriting ``return address'' was also infeasible, since the storage location of function return address is not known.

The second attack is demonstrated on a dynamically allocated object. They observe that addresses of dynamically allocated memory blocks are predictable and address of the virtual table (vtable), which contains addresses of virtual functions defined in the class, can be easily obtained from a dynamically allocated object. By utilizing this, the vtable address of the object can be overwritten. Using this, a malicious function can be called. Although the attacks discussed by them do not pose an explicit threat, their work highlights the possible loopholes in GPU software.

Di et al. \cite{di2016study} study buffer overflow vulnerabilities in CUDA-based GPUs.  By causing a stack overflow, the address of a function pointer can be changed. Using this, execution can be directed to a malicious function. They illustrate this attack in CUDA and confirm that current GPUs do not have a mechanism to mitigate stack overflow. By causing overflow in GlM, the attack launched from a thread can be targeted to other threads. 

As for heap,  since heap memory of different kernels/threads is allocated in contiguous locations, an overflow in the local buffer of a thread can corrupt the heap of another thread. Also, for a globally-visible heap pointer, the memory pointed by it can be accessed by threads which did not originally allocate it. This shows lack of proper access-control in GPUs. Since heap-pointer addresses can be guessed, an adversary can easily launch an attack. Similarly, two kernels that run simultaneously can corrupt the heap memory of each other. Also, for two kernels that run sequentially,  the data of the first kernel remains unchanged in memory and since the pointer-address assigned to the second kernel is same as that of the first kernel, the first kernel can force the second kernel to access garbage data. Further, under/over-flows in integer arithmetic operations and function pointer overflows in {\tt struct} are also possible in CUDA. However, due to features of CUDA, exception handling and format string vulnerabilities cannot be exploited.

\begin{figure} [htbp]
\centering
\includegraphics[scale=0.35]{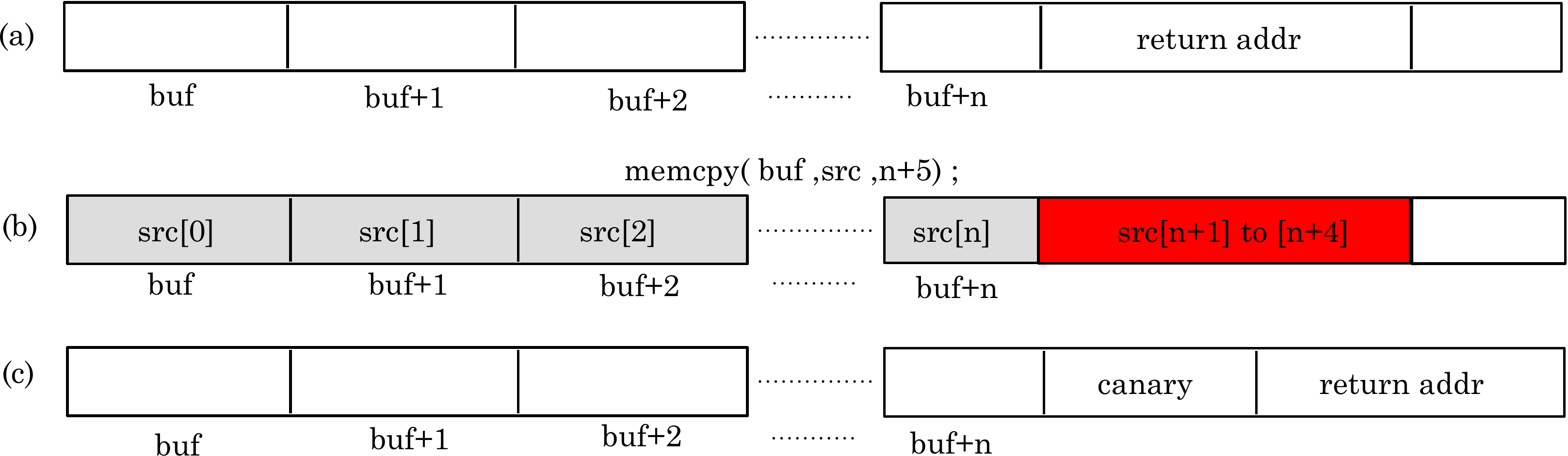}
\caption{ (a) An example of a buffer with indices 0 to {\tt n}. After the buffer, return address of a function is stored. (b) An adversary can copy an excessive amount of data to cause overflow and thus, overwrite nearby variables (return address). (c) Use of a canary value after the buffer \cite{erb2017dynamic}. A change in canary indicates a buffer overflow. }\label{fig:BufferOverflow}
\end{figure}

Erb et al. \cite{erb2017dynamic} present a technique for detecting runtime buffer overflows in OpenCL GPU applications. Figure \ref{fig:BufferOverflow}(a) shows a snapshot of memory, where the return address is located right after the buffer. Figure \ref{fig:BufferOverflow}(b) illustrates overwriting of the return-address by copying excessive amount of data. Their tool adds canary regions outside buffers, as shown in Figure \ref{fig:BufferOverflow}(c). After kernel execution, the canary region is checked to see if the kernel wrote beyond the limits of its memory regions. Any buffer overflow is reported to the user. The tool wraps OpenCL API calls using {\tt LD\_PRELOAD} mechanism. It monitors overflows in global \textit{cl\_mem}, coarse-grained and fine-grained shared virtual memory (SVM) buffers by expanding requested buffer size to include canary regions. It does the same for global \textit{cl\_mem} images by expanding each dimension of an image with canary regions and sub-buffers by creating shadow copies.

In cases where the previously allocated CPU memory is allowed to be used as a buffer, a shadow copy is used to avoid the arduous task of updating the users' pointers. However, for images, the canaries are read into a one-dimensional array, which in turn is fed into a checker along with a buffer containing the end point for each image. SVM buffer overflows are detected using a single buffer containing pointers to the beginning of the canaries in all the SVM buffers.

They note that when the number of buffers is small, the data transfer size is also small. Hence, it is preferable to perform checking of {\tt cl\_mem} buffers on CPU since the kernel invocation overhead of GPU cannot be amortized. However, with increasing number of buffers, a single GPU routine can check multiple canary regions in parallel and hence, using GPU checker is preferable. For SVM buffers and images, CPU checking always incurs higher overhead due to the need for performing data transfers. They show that their tool can detect even those overflows, which are not detected by other tools.  

A limitation of their technique is that canaries are checked only after completion of the kernel.  Hence, before the canaries are checked, a garbage value may get consumed, and the canaries can be reset which avoids detection of overflow.  Further, an adversary who knows the canary size can overwrite beyond it.  Thus, their tool does not fully guarantee security. Also, the use of canaries wastes memory capacity.

\subsection{Denial-of-service attack}\label{sec:denialofservice}

Patterson \cite{patterson2013vulnerability} presents ways to launch DoS attacks   on GPUs. To launch a DoS attack, the GPU  should be assigned a long-running task, which makes it unavailable for other system-tasks such as drawing the desktop. Since a GPU task cannot be pre-empted, the system becomes unresponsive. DoS attacks can be launched using graphics APIs, eg., DirectX,  OpenGL and WebGL. Using WebGL, they show a DoS attack   on the drawing function ({\tt gl.draw}), vertex and fragment shaders. To attack the drawing function, instead of calling it multiple times, it should be called only once and given a large amount of workload, e.g., many complex shapes. This is because, after each invocation of the drawing function, the control returns to CPU, which would foil the DoS attack. Further, the number of shapes should be small enough to fit in the GPU memory and avoid a GPU-crash, and large enough to make GPU unresponsive while rendering them. Similarly, to launch DoS attack on vertex and fragment shaders, an infinite loop can be written in their codes. Table \ref{Table:DoSAttacks}  shows the results of these attacks on various operating systems.

\begin{table}[htbp]
\centering
\caption{Results of DoS attacks on different GPUs and operating systems \cite{patterson2013vulnerability}}
\begin{tabular}{ |l|p{4cm}|p{5.5cm}|p{4cm}| } \hline
  & \multicolumn{1}{c}{Nvidia GPU} &  \multicolumn{1}{|c}{ATI GPU} &  \multicolumn{1}{|c|}{Intel GPU} \\
 \hline

 \multicolumn{4}{|c|}{\textbf{Results of flooding the  {\tt gl.draw} function}} \\\hline

 Windows XP &  Total system freeze & System freeze, then GPU recovery message & Not tested\\
 \hline
 Windows 7 & System freeze, then graphics driver reset & System freeze, then graphics driver reset. Occasional total system freeze. & System freeze, then graphics driver reset.\\
 \hline
 Mac OS X & Total system freeze & Total system freeze & Not tested\\
 \hline
  Red Hat Linux & \multicolumn{3}{G|}{System freeze, then graphics driver reset} \\\hline
\multicolumn{4}{|c|}{\textbf{Results of the attack on vertex/fragment shader}} \\\hline
 Windows XP &  Total system freeze & System freeze, then GPU recovery message & Not tested\\
  \hline  
Windows 7 & \multicolumn{3}{G|}{System freeze, then graphics driver reset} \\
    \hline
 Mac OS X & Total system freeze & Total system freeze & Not tested\\
 \hline
 Red Hat Linux & \multicolumn{3}{c|}{Vertex shader: total system freeze. Fragment shader:  hang, then graphics driver reset}  \\
 \hline

\end{tabular}
\label{Table:DoSAttacks}
\end{table}

As for countermeasures, they note that some OSes use timers to ascertain  when the GPU stops responding and then reset the driver to reclaim GPU. However, they observe that the implementations of these drivers are not perfect and may lead to system crash, e.g., after five timeouts, Windows system crashed completely. Linux OS could not detect the attack on vertex shader and Mac OS X was found to have no protection against this attack.  They also recommend some mitigation strategies for future GPU systems. First, by performing static analysis, the runtime can be estimated and if it exceeds a threshold, the kernel launch can be prohibited. Second, by allowing simultaneous execution of multiple tasks on GPUs, some resources can be ensured for OS processes, even if other task stops responding. Finally, resetting should be performed in time to avoid complete system crash.

\section{Conclusion and Future Outlook}\label{sec:conclusion}

As an increasing number of security vulnerabilities of GPUs come to light, both vendors and users should now carefully weigh the performance advantages of GPUs vis-a-vis their security loopholes. In this paper, we presented a survey of techniques for improving the security of GPUs. We classified the techniques based on key parameters and underscored their similarities and differences. We foresee huge research efforts in the coming years in the area of  both attacking and securing GPUs. We conclude this paper with a discussion of future research directions. 

In the highly-competitive market of today, the choice of computing device depends on many metrics, such as its performance, energy efficiency, security and ease-of-use. However, these metrics are often at odds with each other. Evidently, performance-inefficient security solutions may make GPU an unattractive platform for hardware acceleration of compute-intensive applications. To justify their adoption, future security solutions ideally need to be ``invisible'' from the perspective of their performance/power impact. Alternatively, they should provide a tunable knob to the user to exercise a trade-off between the level of security and acceptable performance loss. Also, by use of novel approaches such as approximate computing \cite{mittal2016SurveyApprox}, the performance overhead of security solutions can be greatly mitigated.  

To enable closer interaction between CPUs and GPUs, recent research has proposed fused chips where a CPU and GPU are integrated on the same chip \cite{mittal2015cpugpusurvey}. It is unclear whether the insights gained for discrete GPUs can be applied for fused GPUs also. Moving forward, an in-depth exploration of the security issues of fused GPUs is definitely required.

We believe that  the challenges involved in ensuring the security of GPUs need to be addressed at all levels of the computing stack. At the hardware level, memories with inherent security properties (e.g., efficiently erasing data at short-notice) need to be used. At the (micro) architecture level, strong encryption algorithms and randomization schemes need to be developed \cite{mittal2018SurveyNVMSecurity}. Also, if future GPUs allow multiple processes to run concurrently, they can   also implement virtual memory. Using this, ASLR and process isolation can be easily implemented. 

At the system level, techniques for mitigating access-violation are required to effectively share GPUs between users without fearing its security implications. Also, the OS should monitor GPU programs and resource usage to detect anomalous behavior. At the application level, GPU library routines should automatically clear memory right after deallocation. Also, the vendors should provide detailed documentation on commercial GPUs and more robust tools for analyzing GPU binaries \cite{albassam2016enforcing}. The standards committee should mandate a thorough security analysis of the entire GPU stack.  We look forward to an exciting future where GPUs pass the scrutiny on the metric of security also, just as they have previously passed the scrutiny on performance/power metrics.

The task of securing GPUs is a never-ending one since, while some researchers design a secure GPU or propose a security technique, other researchers point out its vulnerabilities. Since even one loophole in security can be exploited to take full-control of the system, the goal of security requires the architects to be always on vigil. Clearly, concerted and ongoing efforts are required from both industry and academia to design fully secure GPUs of the next-generation.


\ifCLASSOPTIONcaptionsoff
  \newpage
\fi



%
{\footnotesize
\bibliographystyle{IEEEtran1}
\bibliography{References}
}

%



\end{document}